\documentclass[12pt,preprint]{aastex}

\begin{document}

\title{The Orientation of Jets Relative to Dust Disks in Radio Galaxies\altaffilmark{1}}

\author{H. R. Schmitt\altaffilmark{2,3}, J. E. Pringle,\altaffilmark{4,5},
C. J. Clarke\altaffilmark{4}, A. L. Kinney\altaffilmark{6}}
\altaffiltext{1}{Based partly on observations made with the NASA/ESA Hubble
Space Telescope, which is operated by the Association of Universities for
Research in Astronomy, Inc., under NASA contract NAS 5-26555.}
\altaffiltext{2}{National Radio Astronomy Observatory, P.O. Box O, 1003
Lopezville Road, Socorro, NM 87801.}
\altaffiltext{3}{Jansky Fellow.}
\altaffiltext{4}{Institute of Astronomy, University of Cambridge, Madingley
Road, Cambridge CB3 0HA, UK.}
\altaffiltext{5}{Space Telescope Science Institute, 3700 San Martin Drive, 
Baltimore, MD 21218.}
\altaffiltext{6}{NASA Headquarters,300 E Street, Washington, DC20546.}

\date{\today}

\begin{abstract}

We use HST broad band images, VLA and VLBI continuum data to study the
three dimensional orientation of jets relative to nuclear dust disks
in 20 radio galaxies. The comparison between the position angles of the
jets with those of the dust disks major axes shows a wide distribution,
suggesting that they are not aligned preferentially perpendicular to each
other. We use a statistical technique to determine the 3 dimensional
distribution of angles between jets and dust disks rotation axes. This
analysis shows that the observations are consistent with jets homogeneously
distributed over a large region, extending over polar caps of 55$^{\circ}$
to 77$^{\circ}$, but seem to avoid lying close to the plane of the dust
disks. We argue that the lack of close alignment between jets and dust
disks axes is not likely to be caused by feeding the nucleus with gas from
mergers originated from random directions. We suggest that the misalignment
can be due by a warping mechanism in the accretion disk, like self-irradiation
instability or the Bardeen-Petterson effect, or that the gravitational
potential in the inner regions of the galaxy is misaligned with respect to
that of the dust disk.

\end{abstract}

\keywords{accretion -- accretion disks -- galaxies:active -- galaxies:jets --
galaxies:structure -- galaxies:nuclei}

\section {Introduction}

The orientation of a black hole rotation axis relative to its host galaxy
rotation axis is an important diagnostic for the structure of the 
accretion disk and the nuclear region of the galaxy. In the case of spiral
galaxies, the simplest picture suggests that they should be aligned, since
the gas which fuels the nucleus is expected to originate in the galaxy
disk, or if new gas is added to the galaxy, it will end up in the disk.
However, Clarke, Kinney \& Pringle (1998), Nagar \& Wilson (1999) and Kinney
et al. (2000, hereafter K00), found that this assumption is wrong for
Seyfert galaxies. By comparing the orientation of their radio jets, which
are believed to be perpendicular to accretion disks and aligned with
the black hole axes, with the major axis of their host galaxies, they showed
that the observed distribution can be reproduced by randomly oriented jets.

In the case of radio galaxies the situation is a little different from that of
Seyferts. Most of their hosts are ellipticals or S0s and have very little
cool gas to feed the nucleus, meaning that the gas is likely to have an
external origin, probably from mergers. Birkinshaw \& Davies (1985) and
Sansom et al. (1987) presented the comparison between the position angle
of jets and host galaxy major axes for this class of objects. They found
that there is very little or no correlation between the two, a result similar
to the one found for Seyfert galaxies. This result is contrasted by the one
obtained by Kotanyi \& Ekers (1979), who approached the problem using
dust lanes to trace the orientation of the gas in these galaxies. They found
that radio galaxies have jets aligned nearly perpendicular to these dust
lanes, implying a close connection between the gas in these structures and
the accretion disk. M\"ollenhoff, Hummel \& Bender (1992) obtained a similar
result using ground based data. More recently, van Dokkum \& Franx (1995) and
de Koff et al. (2000) used HST images to detect dust lanes in radio galaxies.
They found a large range of values for the angle between the jets and dust
lanes, but suggest that the majority of the jets are still aligned nearly
perpendicular to the dust lanes (however, see de Ruiter et al. 2001 for a
different result).

Although the previous studies about the orientation of jets in radio galaxies
presented interesting results, they faced several problems. For those papers
which compared the orientation of jets relative to host galaxies major
axes, it would be important to determine if their host galaxies are triaxial
or oblate systems. It was shown by Binney (1985) that the intrinsic major
axis does not necessarily correspond to the projected one in triaxial galaxies,
and consequently the results obtained by Sansom et al. (1987) and Birkinshaw \&
Davies (1985) could be contaminated to orientation effects. In the case
of papers which used the orientation of jets relative to dust lanes, it is
assumed that the gas and dust define a preferential plane, thus eliminating
the uncertainties of host galaxy projection (Kotanyi \& Ekers 1979). However,
the inspection of the HST images presented by De Koff et al. (2000) and Capetti
et al. (2000) shows that most of the dust lanes in radio galaxies are patchy
and not very well organized, suggesting that they may not be dynamically
relaxed systems, thus not directly related to the feeding
of the nucleus. It is extremely difficult, and even impossible in some of
the cases, to determine the orientation and measure a position angle of the
dust in most of these galaxies.

In this letter we address the question of the orientation of jets relative
to dust structures in radio galaxies using a new approach. Instead of using
all radio galaxies with dust lanes, we use only those with well defined
dust disks in their nuclei. We also improve on previous analyzes by using the
statistical techniques devised by K00 (see below).

\section{Data and Measurements}

In order to avoid the uncertainties faced by previous works, we decided to use
only radio galaxies which present a nuclear dust disk. These disks are relaxed
structures, settled in the gravitational potential of the galaxy, for which we
can determine the position angle of the major axis, as well as the inclination
relative to the line of sight, under the assumption that they are circular.
Another advantage of using only dust disks is the fact that it allows us to
use the same technique developed by K00 for Seyfert galaxies and obtain
information about the 3-dimensional distribution of the jets. This technique
determines, in a statistical way, which distributions of angles $\beta$, the
angle between the jet and dust disk rotation axis in 3 dimensions,
are the best representations to the data. Most of the previous studies in this
field used only the difference between the position angles of dust lanes and
jets for their analysis, which are 2-dimensional projections on the plane
of the sky and do not give information about the 3 dimensional distribution.

Our sample was selected from a list of radio galaxies with z$\leq0.1$
available in NED. A source was considered a radio galaxy if its host 
morphology is elliptical or S0, with optical absolute magnitude M$_R\leq-21$
and radio power P$_{1.4GHz}\geq10^{22}$W Hz$^{-1}$. This list was cross
correlated with the HST archive, and the galaxies for which we could find
broad band images were inspected for dust disks. We found a total of
20 galaxies with dust disks which can be used for our analysis. Notice
that this sample is not complete, like the 3C sample, since it involves
galaxies from different surveys (3C, 4C, Bologna, PKS), discovered based
on observations at different frequencies, with different
sensitivities, but it includes all radio galaxies with known dust disks.
If we had used only the galaxies from the 3C survey, the sample would be
reduced to too small a number to allow us to perform our analysis. Of the 20
galaxies in the sample, 13 already had measurements of disk major axis position
angle and inclination published by other authors. For the remaining galaxies
we carried out these measurements by overplotting different ellipses on top of
the images, centered at the nucleus, and determined by eye which one gave the
best reproduction to the observed dust disk. The uncertainties in these
measurements are of the order of 5$^{\circ}$, similar to the ones involved
in the measurements of the position angles of the jets. The smallest scale
on which the dust disks can be traced is 0.1$^{\prime\prime}$, the HST
resolution, which corresponds to sizes between 20~pc and 150~pc, depending
on the distance of the galaxy (we use H$_0=75$~km~s$^{-1}$~Mpc$^{-1}$).
The radio measurements from VLA, and in some cases also from VLBI, were
obtained from the literature, measured directly on published radio images,
and, in some cases, measured on images obtained reducing VLA archival data.

Galaxies like 3C\,40 were excluded from the sample,
because, even though we could see a dust disk, the galaxy does not present
a jet, only the hotspots far away from the nucleus, making it impossible to
obtain a robust estimate of the position angle of the extended radio emission.
Galaxies like Centaurus A and NGC7626 were not included in our study
because their dust disks are warped and we are not able to obtain a reliable
value for their inclinations.
Similarly, cases like 3C\,264, which show a face-on dust disk (Sparks et al.
2000), also had to be eliminated from the sample, since in these cases it is
impossible to constrain the position of the jet relative to the disk.
The fact that Sparks et al. (2000) find round disks supports our assumption
that they are intrinsically round, and will be discussed in detail below.

In Table 1 we present the sample and the measurements. The first column gives
the names of the galaxies, followed by their radial velocities in column 2.
Columns 3, 4, 5 and 6 present the position angles of the dust disks major
axes, the host galaxies major axes, the jets on VLBI scales and on the VLA
scales, respectively. Column 7 presents $\delta$, the difference between the
position angle of the jets measured on VLA images and the dust disks major
axes, column 8 gives the inclination ($i$) of the dust disks relative to the
line of sight and column 9 gives the minimum angle between the jets and the
dust disks rotation axes ($\beta_{min}$).
The morphological types of the host galaxies are given in column 10,
and the references from which the measurements were obtained are presented
in column 11.

\section{Results}

We first compare the orientation of the dust disks relative to their
host galaxies. Figure 1 presents the difference between the position angle of
the host galaxies major axes and dust disks major axes for our sample. In most
of the cases the dust disk major axis is closely aligned with the host galaxy
major axis, with the exception of 3C\,402N and 4C--03.43, where they are
aligned close to the minor axis. Analyzing these two galaxies in detail
we find that 3C\,402N is interacting with other galaxies (de Juan et al.
1994), and the isophotes close to the nucleus change position angle by
90$^{\circ}$ relative to the outer ones, so the dust disk is in fact aligned
with the major axis in the nuclear region. In the case of 4C--03.43, we find
that the host galaxy is almost face-on which makes it very difficult to
accurately measure the position angle of the major axis.

A naive analysis of the result from Figure 1,
ignoring projection effects, might suggest
that these are triaxial galaxies, since according to  Schwarzschild (1979)
only orbits around the major and minor axes are stable in these systems.
However, it was shown by Binney (1985), Franx, Illingworth \& de Zeeuw
(1991) and Ryden (1992) that, due to projection effects, the observed major
and minor axes, projected on the plane of the sky, do not correspond to the
intrinsic ones in triaxial galaxies. In contrast, the distribution seen in
Figure 1 is very narrow, with most of the dust disks aligned very  close to
the  host galaxy major axis, similar to the expected distribution for oblate
systems. This result indicates that our galaxies either have disks or are
oblate axisymmetric systems. In fact, 30\% of them are classified as S0 or
E/SO (Table 1), indicating that they have disks. The remaining galaxies are
classified as ellipticals and we consider them to be oblate systems.
Based on this fact we assume that the dust disks are intrinsically
circular, and use this assumption to derive their inclinations relative to
the line of sight. This agrees with the results from Sparks et al. (2000)
who present several radio galaxies with circular dust disk, seen face-on.

The above result is similar to the one obtained by Tran et al. (2001)
for normal elliptical galaxies, where they also found that their nuclear
dust disks are aligned close to the host galaxies major axes, indicating
that they are oblate systems. A complicating factor in the assumption that
the dust disks are intrinsically circular have been recently presented by
Andersen \& Bershady (2002), who showed that disk galaxies
are not always exactly round, but have slightly elongated shapes, with
an average ellipticity e$=0.06$. If we assume that the intrinsic ellipticity
of all our dust disks is e$=0.06$, we get that in 75\% of the cases, those
with $i\geq60^{\circ}$, the inclinations would be reduced by
$\sim1^{\circ}-2^{\circ}$, which is smaller than
the uncertainty in the measurements.

Figure 2 presents the distribution of $\delta$'s, obtained using the jet
directions measured on VLA scales. We prefer to use the measurements based
on VLA images because they are available for all galaxies in the sample.
Although the VLBI measurements would trace the jets in regions much closer
to the nucleus, they are available only for 8 galaxies. The position angles
obtained from VLBI differ by no more than 7$^{\circ}$ from the VLA results,
and making use of the VLBI measurements does not significantly affect our
conclusions. The inspection of this Figure shows that there is a large
range of $\delta$'s in our sample from 19$^{\circ}$ to 89$^{\circ}$,
suggesting that there is not a good alignment of the jets perpendicular to
the dust disks. As explained above, the $\delta$ distribution is a
2-dimensional projection on the plane of the sky, and we are actually
interested on the orientation of jets in 3-dimensions, distribution of
angles $\beta$, which is obtained using the techniques developed by K00.

We give here a brief description of the geometry of the
problem, and a more detailed description can be seen in K00. We assume the
nucleus to be centered at the origin of the reference frame, with the
dust disk lying in the XY-plane, the major axis along the X-axis,
the rotation axis along the Z-axis, and the dust disk inclined relative to
the line of sight by an angle $i$. The dust disk major axis and the jet are
projected onto the sky plane, and the difference between their position
angles is $\delta$. The angle between the jet and the galaxy rotation
axis is $\beta$. It was shown by Clarke et al. (1998) that for a given galaxy
with a measured pair of angles $i$ and $\delta$, the jet is determined to lie
on two great circles, which are mirror reflections of each other. These great
circles are drawn on a sphere centered at the origin of the system, the
nucleus. A particular property of the geometry of this problem is the fact
that all points in a great circle lie at angles $\beta\geq\beta_{min}$, meaning
that for a given pair ($i$,$\delta$) the jet will never be at an angle closer
than $\beta_{min} = \sin^{-1} ( \sin i \cos \delta)$ relative to the disk
rotation axis, but can always be as large as 90$^{\circ}$, along the equator.
In order words, for a dust disk of inclination $i$, there is a large
range of $\beta$ angles which can reproduce an observed value of $\delta$.

The values of $\beta_{min}$
for our sample are presented in Table 1 and the cumulative distribution of
these angles is presented in Figure 3. This distribution varies smoothly and
shows a large range of values. In the case of some radio galaxies, like
NGC\,315, the jet could be very well aligned with the dust disk rotation axis,
with $\beta_{min}=1^{\circ}$. However, in most of the cases the jet has to be
misaligned by at least 20$^{\circ}$, with 30\% of the sample having
$\beta_{min}>30^{\circ}$. The largest value of $\beta_{min}$ corresponds to
53$^{\circ}$ for UGC\,12064. The uncertainties in the values of $\beta_{min}$,
calculated using propagation of errors in $i$ and $\delta$,
is of the order of 4$^{\circ}-5^{\circ}$.

We would like to be able to determine the value of $\beta$ for each one of
the galaxies. Unfortunately, with the exception of a small number of cases
(Pringle et al. 1999), we cannot determine this value directly, but rather a
range of acceptable values. The distribution of $\beta_{min}$'s already
indicates that the jets do not lie preferentially
perpendicular to the dust disks. In order to improve on this
result we used the method described by K00 to test whether a homogeneous
distribution of jets over polar caps of different sizes, in the range
0$^{\circ}\leq\beta\leq\beta_2$, can represent the observed $\delta$
distribution. This technique assumes a homogeneous distribution of $\beta$'s
and uses, for each one of the galaxies, the information about $i$ and $\delta$.
This distribution is integrated between 0$\leq\delta\leq\delta_k$, where
$\delta_k$ is the observed value of $\delta$ for the galaxy, to obtain the
centile point $c_k$. The final distribution of $c_k$'s is then compared with
a homogeneous distribution in the interval [0,1] using the Kolmogorov-Smirnov
test (see K00 for a detailed explanation of the technique).

We tested which distributions of jets can represent the observed data, by
varying $\beta_2$ between 55$^{\circ}$ and 90$^{\circ}$. The lower limit
(55$^{\circ}$) was chosen based on the fact that the largest value of
$\beta_{min}$ is $53^{\circ}$, so any distribution of jets
with $\beta_2<53^{\circ}$ would not include all the
galaxies available in the sample. We find that polar
caps with $\beta_2$ between $55^{\circ}$ and $77^{\circ}$, which correspond to
regions between 40\% and 78\% of the parameter space, respectively, are good
representations to the data. The KS test shows that the hypothesis that the
observed data is drawn from a homogeneous distribution of jets over a polar cap
with $\beta_2=55^{\circ}$ is consistent at the 94\% level, while a polar cap
with $\beta_2=77^{\circ}$ is consistent at the 5\% level. For polar caps with
$\beta_2>77^{\circ}$ the KS test gives increasingly smaller probabilities of
the models agreeing with the observations, reaching a value of 0.9\% for jets
distributed over the entire sphere ($0^{\circ}\leq\beta\leq90^{\circ}$).
Loosely speaking, our finding is that, if the jets are uniformly distributed
over angles between 0$^{\circ}$ and $\beta_2$ relative to the disk axis, then
at the 2$\sigma$ level $\beta_2$ lies between 55$^{\circ}$ and 77$^{\circ}$.

\section{Discussion}

The above results clearly show that jets are not closely aligned perpendicular
to dust disks. They can be represented by homogeneous distributions over large
polar caps ($55^{\circ}\leq\beta_2\leq77^{\circ}$), encompassing 40\% to 78\%
of the parameter space. The smaller polar caps agree in part with the
suggestion by Sparks et al. (2000) that jets are aligned at
30$^{\circ}-40^{\circ}$ from the rotation axis, but a larger parameter space
cannot be ruled out by the current data. The models show a probability between
5\% and 0.9\% that the jets uniformly populate a region extending more than
$77^{\circ}$ from the pole. This implies a deficit of jets that are closely
aligned along dust disks, as one might expect since they
would either destroy the dust disk, or not be able to propagate through it.

The reason why other papers found a better alignment between dust structures
(lanes and disks) and radio jets is not very clear. Some of the previous
studies were based on small samples (less than 10 galaxies), where selection
effects and small number statistics may have influenced the results.
Something else which might have influenced the results presented by other
papers is the fact that most of them based their conclusions only on the
difference between the position angles of the jets and dust structures.
As mentioned above, these measurements are only the projection of the
intrinsic 3 dimensional distribution on the plane of the sky. They can
have a much tighter distribution than the intrinsic one, creating the
impression of a better alignment perpendicular to the dust lanes.

As is also the case in Seyfert galaxies, the result we obtained contradicts
the simplest expectations that the black hole should have its angular
momentum vector aligned to that of the host galaxy. In their study, K00
presented a series of mechanisms to explain how the observed misalignment
could happen, some of which can be used to explain radio galaxies.
Misaligned inflow of gas towards the nucleus of the galaxy could, in principle,
explain the results. Since we are dealing with early type
galaxies, the origin of gas to feed the nucleus is probably external from the
galaxy, from mergers with gas rich galaxies.
Under this scenario, the observed jet would be the result of a previous merger,
while the observed dust disk is the result of a more recent one, which
had a different orientation, and will produce jets oriented in a different
direction. There are several evidence which point against this scenario.
First, we showed that the dust disks are closely aligned with the host galaxy
major and minor axes, meaning that they had enough time to settle into the
gravitational potential of the galaxy and are not the result of recent
accretion. Second, the comparison between the orientation of jets on small (VLBI)
and large (VLA) scales also indicates that their directions are not regulated
by mergers. For 8 of the galaxies we found high resolution VLBI
continuum data, which show an alignment better than 7$^{\circ}$ with the
VLA scale jets, showing their long stability over time, from 1~pc to
10~kpc scales.

Since the feeding of the nucleus by a misaligned inflow of gas does not seem
to explain the observations, the misalignment has to occur between the
inner region of the dust disk, which corresponds to dimensions between 20~pc 
and 150~pc for the galaxies studied here, and the point where the jet is
produced (a few Schwarzshild radii). This can be achieved by
the warping of the accretion disk by self-irradiation
instability (Pringle 1996, 1997; Maloney, Begelman \& Pringle 1996),
by the Bardeen-Petterson effect in the case
the galaxy has a rapidly rotating black hole, or by a misaligned gravitational
potential in the nuclear region of the galaxy. However, some of these
mechanisms may present some problems in radio galaxies.

Warping of the accretion disk by self-irradiation instability, possibly
enhanced by mhd wind warping, could explain the observed misalignment.
In this case the black hole should not be rotating rapidly, as suggested by
Blandford (1990) to be the case in radio galaxies. According to
Pringle (1997), a rapidly rotating black hole stabilizes the inner region
of the accretion disk and can inhibit the self-irradiation instability.

Warping of the accretion disk by the Bardeen-Petterson effect can explain
misaligned dust disk and jet axes if the black hole is rotating rapidly,
with a rotation axis different from that of the galaxy and the dust disk. This
can result from the merger between two black holes (Wilson \& Colbert 1995),
originated from a galaxy merger, but the time scale should be long enough
to allow the gas and dust to settle in the galaxy potential. It is also
possible that the black hole with misaligned axis relative to the host galaxy
was created by a previous merger, while the gas and dust we see are the result
of a more recent interaction. Here the rapidly rotating black hole forces the
precession and alignment of the inner regions of the accretion disk with its
spin axis. One problem with this scenario, pointed out by Rees (1978) and
more recently by Natarajan \& Pringle (1998), is that the black hole exerts
a torque on the accretion disk, which will also exert a torque on the black
hole, eventually forcing the alignment between the black hole and the source of
gas feeding it, in this case the dust disk. The timescale for this alignment
is somewhat uncertain, but if it is relatively fast it might contradict the
fact that we observe the jets to be aligned from the VLBI to the VLA scales.

The warping of the accretion disk could also be due to a misaligned
gravitational potential around the nucleus. Since black holes seem to be
surrounded by a nuclear star cluster (Gebhardt et al. 1996; Faber et al. 1997;
van der Marel 1999), if the star cluster is axisymmetric and has its symmetry
axis misaligned relative to that of the galaxy, it would make the disk
precess and force its alignment with the symmetry axis of the cluster.
This effect may not be very important in radio galaxies. Considering that
their black holes have masses of 10$^8$ to 10$^9$ M$_{\odot}$, their radius
of influence is of the order of 100~pc. A globular cluster would need
to have a similar radius and be axisymmetric, in order to induce the
precession of the accretion disk.

\acknowledgements We would like to thank J. Ulvestad and R. Antonucci for
helpful comments. JEP is grateful for continued support from the
STScI visitor program. This work was partially supported by the NASA grant
AR-8375.01-97A. The National Radio Astronomy Observatory is a
facility of the National Science Foundation operated under cooperative
agreement by Associated Universities, Inc.
This research made use of the NASA/IPAC Extragalactic Database (NED),
which is operated by the Jep Propulsion Laboratory, Caltech, under
contract with NASA.

\begin{deluxetable}{lrrrrrrrrrr}
\tablewidth{0pc}
\tablecaption{Properties of dust disks, host galaxies and radio jets\tablenotemark{1}}
\tablehead{\colhead{Name}&\colhead{Vel}&\colhead{PA$_d$}&\colhead{PA$_h$}&\colhead{PA$_{VLBI}$}&\colhead{PA$_{VLA}$}&\colhead{$\delta$}&\colhead{$i$}&
\colhead{$\beta_{min}$}&\colhead{T}&\colhead{References}\\
\colhead{}&\colhead{(km~s$^{-1}$)}&\colhead{($^{\circ}$)}&\colhead{($^{\circ}$)}&\colhead{($^{\circ}$)}&\colhead{($^{\circ}$)}&\colhead{($^{\circ}$)}&\colhead{($^{\circ}$)}&\colhead{($^{\circ}$)}&\colhead{}&\colhead{}}
\startdata
NGC\,315      & 4942& 40 &   39&   -50 &   -49&   89&  77&  1 &E &1,3\\
NGC\,383      & 5098&140 &  145&   -15 &   -21&   19&  39&  37&S0&4 \\
NGC\,2329     & 5795&174 &  171&   -30 &   -29&   23&  47&  42&E/S0&1,3,15,10\\
NGC\,3557     & 3067& 31 &   34&   --- &    78&   47&  60&  36&E &12,14,19 \\
NGC\,4261     & 2238&165 &  158&   -93 &   -96&   81&  65&  8 &E &3,4 \\
NGC\,4335     & 4629&158 &  156&   --- &    79&   79&  66&  10&E &1,3,5,10 \\
NGC\,4869     & 6875& -9 &  -11&   --- &   -70&   61&  78&  28&E &12,17 \\
NGC\,5127     & 4782& 48 &   68&   --- &   118&   70&  76&  19&E &1,3,10,16 \\
NGC\,5141     & 5303& 88 &   65&   190 &   192&   76&  76&  14&S0&1,2,3,10 \\
NGC\,5532     & 7116&155 &  146&   --- &    35&   60&  72&  28&S0&4 \\
NGC\,6251     & 7459&  4 &   21&   297 &   294&   70&  76&  19&E &8,9 \\
NGC\,7052     & 4672& 65 &   64&   203 &   202&   43&  70&  43&E &1,2,3 \\
NGC\,7720     & 9060& 10 &   28&   -49 &   -56&   66&  43&  16&E &4 \\
IC\,4296      & 3737& 79 &   60&   --- &   130&   51&  72&  37&E &12,19 \\
MCG\,-02-36-02&10999& 55 &   53&   --- &   -60&   65&  66&  23&S0 &12,19 \\
UGC\,367      & 9548&158 &  150&   --- &   -87&   65&  71&  24&E &2,11,12 \\
UGC\,12064    & 5136&160 &  176&   --- &   189&   29&  66&  53&E/S0 &4 \\
3\,C\,402\,N  & 7603& 55 &  150&   --- &   175&   60&  68&  28&E &5,6,7 \\
4\,C\,-03.43  &15499& 73 &  -25&   --- &    13&   60&  67&  27&E &12,18\\
B\,2 0915+32  &18587&109 &  123&   --- &    30&   79&  28&  5 &E &12,13\\ 
\enddata
\tablenotetext{1}{Column 1: galaxy name; Column 2: Radial Velocity:
Column 3: position angle of the
dust disk major axis; Column 4: position angle of the host galaxy major axis;
Column 5: position angle of the VLBI scale jet; Column 6: position angle of
the jet on VLA scales; Column 7: the difference between the position angle
of the dust disk major axis and jet axis; Column 8: dust disk inclination;
Column 9: minimum value of the $\beta$ angle; Column 10: morphological type;
Column 11: references from which the data was obtained:
1-) Verdoes Kleijn et al. (1999), 2-) Fanti et al. (1986), 3-) Xu et al. (2000),
4-) Martel et al. (2000), 5-) Condon \& Broderick (1988), 6-) de Juan,  Colina,
\& Golombek (1996), 7-) de Juan, Colina \& P\'erez-Fournon (1994),
8-) Ferrarese \& Ford (1999), 9-) Jones et al. (1986), 10-) RC3, 11-)
Capetti et al. (2000), 12-) Our measurements from HST data, 13-) Fanti et al.
(1987), 14-) Birkinshaw \& Davies (1985), 15-) Wrobel \& Heeschen (1984),
16-) Fanti et al. (2000), 17-) measured from
Feretti et al. (1990), 18-) measured from Owen, White \& Burns (1992),
19-) measured from VLA archive images.}
\end{deluxetable}

\begin{figure}
\plotone{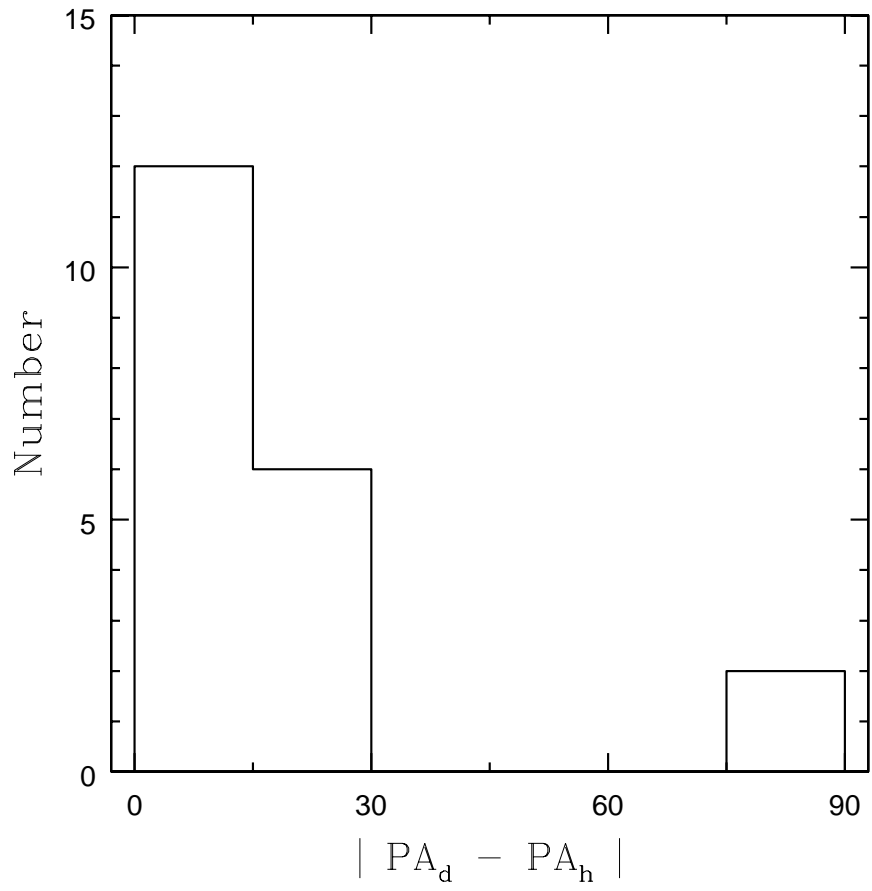}
\caption{Histogram of the difference between the position angle of the
dust disk major axes and host galaxy major axes.}
\end{figure}

\begin{figure}
\plotone{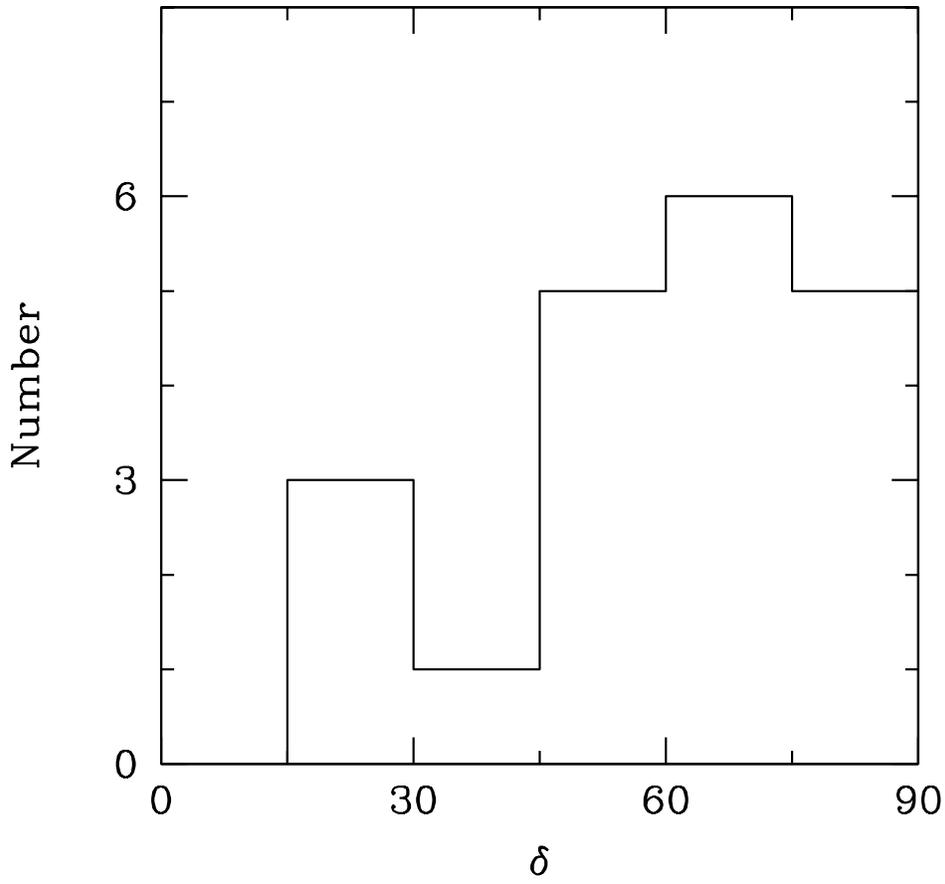}
\caption{Histogram of the difference between the position angle of jets and
dust disk major axes $\delta= | PA_{VLA} - PA_{d} |$}
\end{figure}

\begin{figure}
\plotone{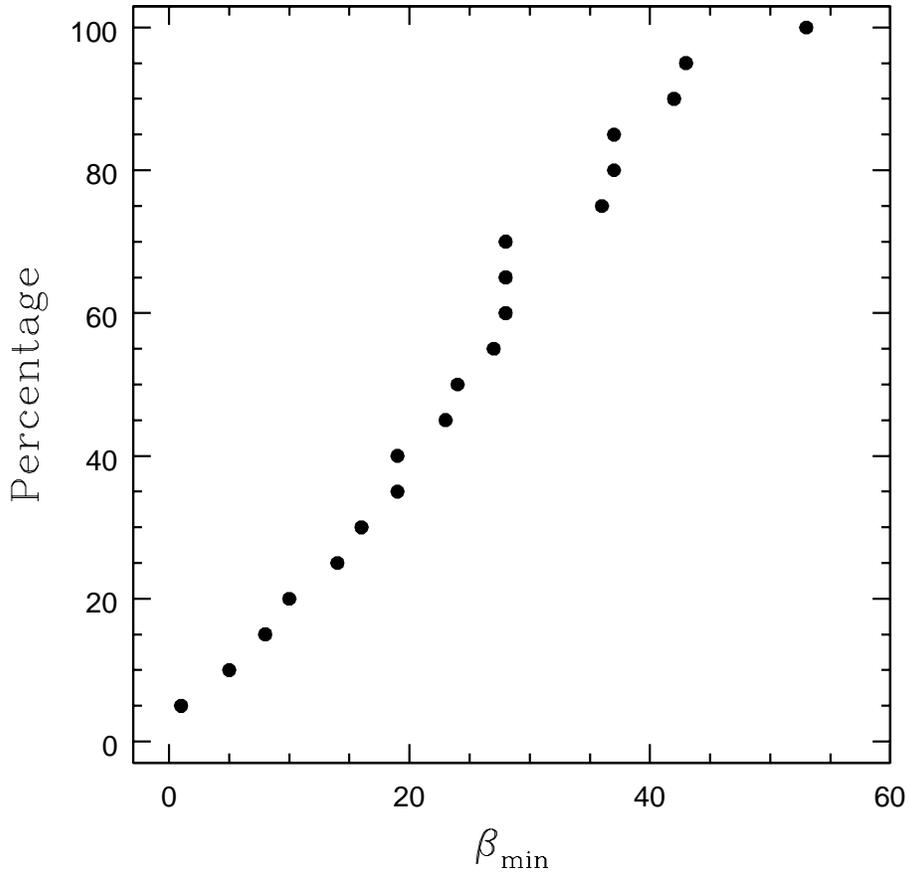}
\caption{Cumulative distribution of $\beta_{min}$ angles, the minimum angle
the radio jet can have relative to the dust disk rotation axis, calculated
using the $i$ and $\delta$ values in Table 1.}
\end{figure}                                           

\end{document}